\documentclass[12pt,a4paper]{article}
\usepackage[utf8]{inputenc}
\usepackage{amsmath,amssymb,amsfonts}
\usepackage{bm}
\usepackage{graphicx}
\usepackage{float}
\usepackage[colorlinks=true,citecolor=blue,linkcolor=blue]{hyperref}
\usepackage{booktabs}
\usepackage{natbib}

\title{Differentiable Particle-Mesh Ewald with Cartesian Tensor Message Passing\\
for Learning Long-Range Electrostatics and Dipole Response}

\author{Zhiyue Guo\textsuperscript{1}, Junjie Wang\textsuperscript{1*}, Haoting Zhang\textsuperscript{1}, Zhixin Liang\textsuperscript{1}\\
Ziyang Yang\textsuperscript{1}, Yujian Pan\textsuperscript{1}, Jian Sun\textsuperscript{1*}\\[0.4em]
\begin{tabular}{c}
\small \textsuperscript{1}National Laboratory of Solid State Microstructures,\\
\small School of Physics and Collaborative Innovation Center\\
\small of Advanced Microstructures,\\
\small Nanjing University, Nanjing 210093, China
\end{tabular}}

\date{}

\begin{document}
\maketitle

\begin{abstract}
Machine learning interatomic potentials (MLIPs) can approach quantum accuracy
for short-range chemistry, yet most architectures remain fundamentally local and fail to
describe the long-range electrostatic and polarization interactions essential for ionic,
polar, and interfacial systems. Recent and contemporaneous Ewald-based MLIPs have
shown that locally predicted electrostatic variables can recover important long-range
physics, including multipolar response. However, many energy-based implementations
still evaluate the reciprocal term by direct sums over $k$ vectors. This leaves a
practical gap between long-range MLIP
development and production molecular dynamics, where particle-mesh Ewald (PME)
with $O(N\log N)$ scaling is the standard.
Here we introduce a fully differentiable PME framework for learned charges and,
centrally, learned atomic dipoles within an E($n$)-equivariant Cartesian tensor
message passing network. Charges are predicted from scalar local features, while
dipole vectors are predicted from equivariant vector features and enter the same
particle-mesh solver as an effective bound charge density. This dipolar density is
constructed by analytic real-space gradients of the Hockney--Eastwood spline
assignment weights, allowing charge--dipole and dipole--dipole long-range forces to
be trained end-to-end through FFT-space electrostatics without direct charge or
dipole supervision.
On a charged-dimer test case, the differentiable PME module reproduces explicit
Ewald energies and forces to numerical precision when the assignment-kernel
deconvolution is enabled. On molten NaCl, the charge+dipole long-range channel gives
the lowest force RMSE among the tested models while all energy RMSE values remain in
the sub-meV/atom regime. Timing tests show the expected crossover from explicit
Ewald summation to particle-mesh scaling. These results
identify differentiable dipole PME as a scalable route toward polarization-aware
MLIPs for condensed-phase and interfacial systems.
\end{abstract}

\section{Introduction}

Machine learning interatomic potentials (MLIPs) have transformed atomistic materials modeling
over the past decade, achieving first-principles accuracy at a fraction of the computational
cost \cite{behler2007,bartok2010,thompson2015,shapeev2016,smith2017ani,
zhang2018dp,schuett2018,batzner2022,batatia2022,wang2024,unke2021}.
By learning the Born--Oppenheimer potential energy surface directly from electronic structure
reference data, these methods have enabled simulations at time and length scales far beyond
the reach of \textit{ab initio} molecular dynamics.

This progress has followed a clear architectural trajectory. Early neural-network and
kernel potentials demonstrated that atom-centered descriptors could learn accurate local
energy decompositions \cite{behler2007,bartok2010,smith2017ani}, while moment and spectral
descriptor families made this idea systematically improvable for materials
\cite{thompson2015,shapeev2016}. Deep message-passing and equivariant architectures then
replaced fixed descriptors with learned geometric representations. These models respect
the fundamental symmetries of atomistic systems: the total energy is invariant to global
translations and permutations of equivalent atoms, atom-wise features are permutation
equivariant with respect to atom indexing, and scalar, vector, and tensor channels
transform consistently under rotations. This symmetry-aware structure improves data
efficiency for energies, forces, and tensorial properties
\cite{schuett2018,schuett2021painn,batzner2022,batatia2022,musaelian2023,
wang2024}.

Despite these advances, most MLIPs still rely on a \textit{locality ansatz}: the total
energy is decomposed into atom-centered contributions that depend only on the atomic
environment within a finite cutoff radius $r_c$. The assumption is physically motivated
by the nearsightedness of electronic matter and gives the linear scaling that makes
large-scale simulations possible. It is also the source of a sharp limitation. Coulomb
interactions decay as $r^{-1}$, charge--dipole interactions as $r^{-2}$, and polarization
response can depend on the dielectric environment over length scales far beyond a
typical 4--6 \AA{} local cutoff. Short-range MLIPs can therefore interpolate accurately
within many covalent and metallic training domains while still failing in ionic
materials, polar molecular liquids, solid electrolytes \cite{you2024llzo},
solid--liquid interfaces, and any system where charge redistribution or long-range
screening controls the forces \cite{prodan2005,anstine2023longrange}.

Existing long-range MLIP strategies can be organized by how they relax this locality
constraint. The most direct route is to enlarge the receptive field, either by increasing
the cutoff, stacking more message-passing layers, or using multiscale short/long-cutoff
modules. This keeps the model architecture close to standard local MLIPs, but the cost
and the number of neighbors grow rapidly, while distant electrostatic information is
still only represented implicitly. A second route adds physically motivated energy
terms. PhysNet and SpookyNet, for example, combine neural short-range energies with
learned charges, dispersion corrections, total-charge or spin information, and damped
electrostatic terms \cite{physnet2019,spookynet2021}. Such decompositions impose useful
asymptotic structure, but the learned electrostatic variables may depend on a chosen
charge convention and are not always sufficient to describe non-local charge
redistribution.

Global electrostatic-variable models address part of this problem by constraining the
charges themselves. Charge-equilibration approaches predict local electronegativities
and hardnesses and then solve a global minimization problem to obtain charges
\cite{mortier1986,rappe1991,streitz1994,behler2015,ko2021}. These methods can capture
non-local charge transfer, but they introduce an additional global solve and inherit the
well-known ambiguities of partial-charge definitions. Deep potential long-range (DPLR)
models avoid direct atomic-charge fitting by using maximally localized Wannier centers
as physically motivated charge carriers \cite{zhang2022dplr}; this provides a smoother
description of electrostatics but requires Wannier information in the reference data.
Classical polarizable force fields offer a complementary lesson: induced dipoles and
atomic multipoles are often essential for dielectric response beyond fixed charges
\cite{thole1981,ren2003amoeba}, but in conventional force fields these variables are
usually parameterized rather than learned end-to-end from quantum-mechanical
energy-force data.

A third family uses global descriptors, reciprocal operators, or non-local
message passing to carry information beyond the local cutoff. Reciprocal-space
neural networks construct structure-factor-like descriptors or learned reciprocal
potentials from the full simulation cell, allowing local MLIPs to access full-cell
information without assigning explicit atomic charges \cite{yu2022rsnn}.
LODE-type descriptors augment local representations with fields generated by
long-ranged density kernels \cite{grisafi2019lode,huguenin2023lode}. Ewald
message passing instead uses reciprocal-space filters as a non-local
message-passing operation \cite{kosmala2023ewaldmp}. Along the same line, LOREM
treats inverse-power-law potential evaluation as equivariant message passing:
scalar and higher-order equivariant charge channels transmit orientation-dependent
geometric information beyond the cutoff \cite{rumiantsev2025lorem}. These latent
channels are not identical to physical atomic multipoles, but they are closely
related in spirit to using high-order tensor representations as carriers of
long-range information.

A complementary energy-based route is to predict electrostatic variables and
evaluate a separate long-range energy. The Latent Ewald Summation (LES) framework
uses local descriptors to predict latent scalar charges and trains them only
through total energies and forces \cite{cheng2023latent,king2025charges},
avoiding direct supervision on convention-dependent partial charges. Kim
et al.\ \cite{kim2026} recently extended this strategy to a polarizable multipole
hierarchy, including monopoles, dipoles, quadrupoles, and non-self-consistent
linear response. Together, these developments show a broad movement toward
combining local equivariant representations with global electrostatic or
reciprocal-space operators. They differ in what is made non-local---descriptors,
messages, latent charges, or multipoles---and should therefore be viewed as
complementary rather than as a single settled formulation.

The specific gap addressed here is the numerical realization of an energy-based
learned charge--dipole branch for periodic, large-scale force-field training. A
direct Ewald implementation evaluates the reciprocal-space contribution as a sum
over all integer vectors $\mathbf{n}$ within a cutoff sphere in $k$-space
\cite{ewald1921}. For a system of $N$ atoms, the number of reciprocal-space
vectors grows superlinearly, yielding an effective complexity of $O(N^{3/2})$ or
$O(N^2)$ depending on the choice of Ewald splitting parameter \cite{kolafa1992}.
This is accurate and useful for reference calculations, but it contrasts with
the state of practice in classical molecular dynamics: major production MD
codes---AMBER, GROMACS, LAMMPS, NAMD, CHARMM---evaluate long-range electrostatics
using the particle-mesh Ewald (PME) method
\cite{case2005,brooks2009,abraham2015,phillips2020,thompson2022lammps,darden1993,
essmann1995}, which accelerates reciprocal-space evaluation to $O(N\log N)$
through the use of the fast Fourier transform (FFT) on a regular mesh. For
atomistic ML, Loche et al.\ recently exposed the torch-pme library \cite{loche2025}. The charge-only PME branch used here was developed with
reference to this framework, while the formulation introduced below
additionally handles learned dipolar densities in a unified differentiable mesh
pipeline. For
learned electrostatic energy models that rely on direct reciprocal-vector sums,
replacing the sum by a differentiable particle-mesh operation is therefore an
important step toward deployable large-scale simulations.

In this work, we bridge this gap by introducing a fully differentiable particle-mesh
Ewald framework integrated with an E($n$)-equivariant Cartesian tensor message passing
architecture~\cite{wang2024}. The long-range solver is implemented as a native
PyTorch module, so gradients propagate end-to-end from the electrostatic energy
through mesh assignment, FFT-based reciprocal-space evaluation, and back to the
latent charge and dipole readouts while retaining the $O(N\log N)$ scaling of
PME.

We examine the framework on two completed example systems. The first is a
charged-dimer test case based on the charged molecular dimer data discussed by
King et al.~\cite{king2025charges}, where long-range electrostatics dominate
the interaction and PME can be compared directly against an explicit Ewald
reference. The second uses liquid NaCl structures from the density-based
long-range descriptor study of Faller et al.~\cite{faller2024density}, which
was also used as a molten-salt example in the LES work~\cite{cheng2023latent}.
Replicated NaCl supercells are used strictly for timing. These examples establish
the two claims supported by the present data: long-range charge/dipole channels
improve the learned force field in systems with strong electrostatics, and the
differentiable PME implementation recovers Ewald-level reciprocal-space accuracy
at particle-mesh cost.
The total energy decomposes as
$E_{\mathrm{total}} = \sum_i E_i^{\mathrm{sr}} + E^{\mathrm{lr}}(\{q_i\}, \{\bm{\mu}_i\})$,
where the short-range term is predicted by the equivariant message-passing network and the
long-range term is evaluated through the differentiable PME pipeline with its gradient
flowing end-to-end back to the network parameters.
In this formulation, PME is not only an efficient electrostatic solver but also a
trainable long-range layer that allows charge--dipole physics to be learned together
with the short-range potential.

\section{Theory and Methodology}

\begin{figure}[t]
\centering
\includegraphics[width=0.98\textwidth]{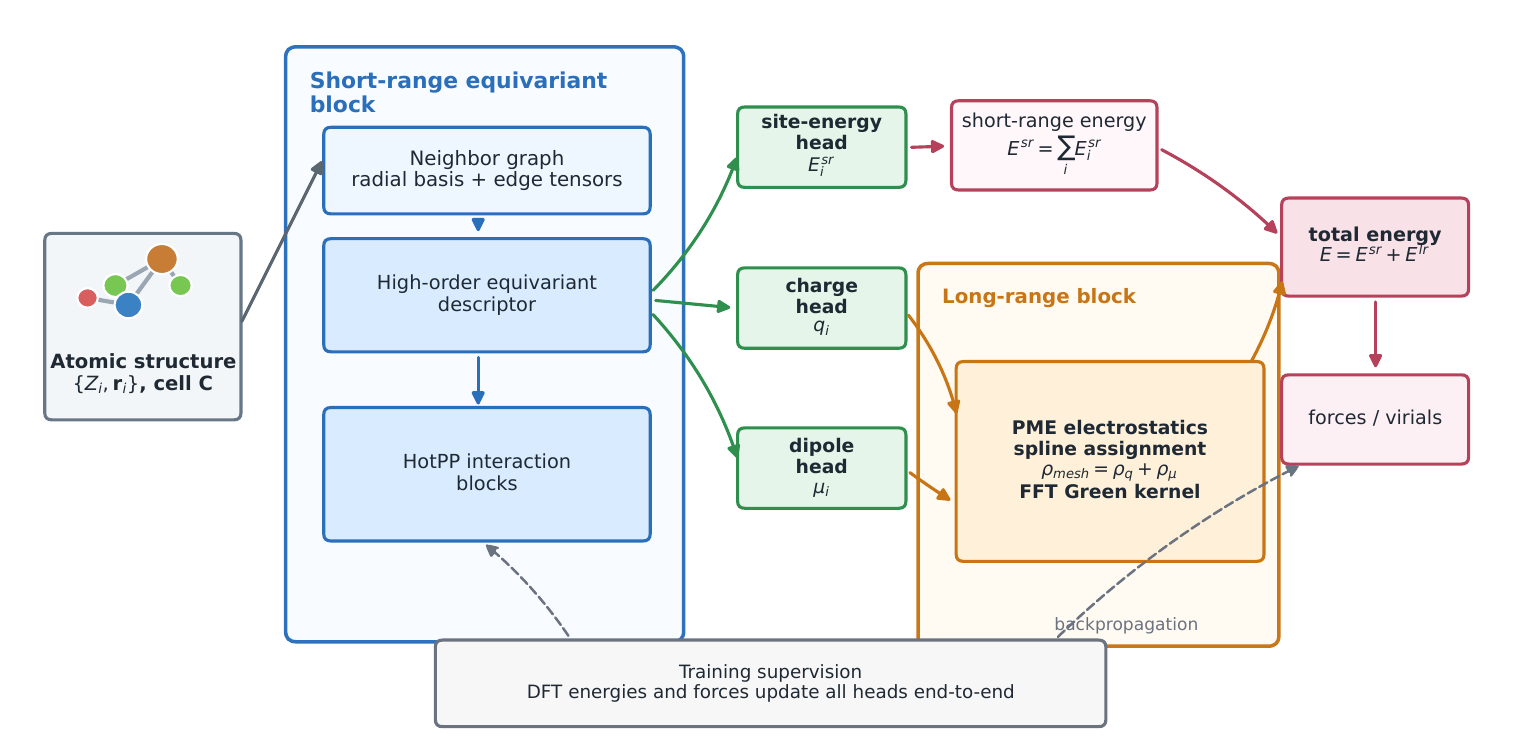}
\caption{Schematic architecture of the HotPP-LR model. The short-range branch is
the E($n$)-equivariant HotPP Cartesian tensor message-passing network. Scalar
and vector features feed separate site-energy, charge, and dipole readouts. The
site-energy head gives $E^{\mathrm{sr}}$, while the charge and dipole heads feed
a differentiable PME branch that constructs $\rho_{\mathrm{mesh}}$, applies the
FFT-space Green kernel, and returns
$E^{\mathrm{lr}}$. The sum $E^{\mathrm{sr}}+E^{\mathrm{lr}}$ is differentiated
end-to-end to obtain forces and train all readout heads from energy-force data.}
\label{fig:model_architecture}
\end{figure}

The overall HotPP-LR architecture is summarized in
Fig.~\ref{fig:model_architecture}. Atomic structures are first encoded by the
short-range HotPP Cartesian tensor message-passing backbone. Separate readout
heads then produce local site energies, latent atomic charges, and latent
atomic dipoles. The site energies define the short-range contribution, whereas
the learned charges and dipoles are deposited onto the PME mesh to evaluate the
non-local electrostatic contribution. The total energy remains differentiable
with respect to positions and network parameters, so force training can update
both the short-range backbone and the long-range electrostatic readouts.

\subsection{Cartesian tensor message passing}\label{sec:hotpp}

We adopt the E($n$)-equivariant Cartesian tensor message passing framework introduced in
Ref.~\citenum{wang2024} as the short-range backbone of our model. Because the
architecture and its tensor algebra are described in detail in the original HotPP
paper, we only summarize the ingredients needed to define the long-range
extension.

For each atom, HotPP stores local features as Cartesian tensors of different
ranks. The scalar rank, denoted here as $\mathbf{h}_i^{(t,0)}$, carries
rotation-invariant chemical information; the vector rank
$\mathbf{h}_i^{(t,1)}$ carries directional information; and higher-rank tensor
features can encode more complex angular response. Message-passing layers update
these features by combining radial basis functions with geometric edge tensors,
followed by self-interaction, nonlinear activation, and residual mixing. This
gives an E($n$)-equivariant local representation after $T$ layers while avoiding
spherical-harmonic basis transformations.

In the present work, we keep this HotPP backbone unchanged and use it to provide
the local, short-range part of the potential. Our contribution begins at the
readout level: the same final local features that predict the short-range site
energy are also used to predict latent charges and dipoles, which are then passed
to a differentiable PME solver.

\subsection{Predicting charges and dipoles from local features}\label{sec:readout}

A key advantage of this Cartesian tensor representation is that different physical
observables naturally emerge from different ranks of the node features. We attach
a multi-head readout layer after the final message-passing layer to predict:

\begin{itemize}
    \item \textbf{Atomic partial charges} $q_i$ from the scalar ($L=0$) features:
    \begin{equation}
        q_i = \mathrm{MLP}_q(\mathbf{h}_i^{(T,0)}),
        \label{eq:charge_pred}
    \end{equation}
    where $\mathrm{MLP}_q$ is a two-layer perceptron with SiLU activation.

    \item \textbf{Atomic dipole vectors} $\bm{\mu}_i$ from the vector ($L=1$) features:
    \begin{equation}
        \bm{\mu}_i = \mathrm{MLP}_\mu(\mathbf{h}_i^{(T,1)}),
        \label{eq:dipole_pred}
    \end{equation}
    where $\mathrm{MLP}_\mu$ acts on the channel dimension, preserving the Cartesian
    vector index. The output is a 3-dimensional vector $\bm{\mu}_i = (\mu_i^x, \mu_i^y, \mu_i^z)$
    that transforms covariantly under rotations.

    \item \textbf{Short-range site energy} $E_i^{\mathrm{sr}}$ from the scalar features:
    \begin{equation}
        E_i^{\mathrm{sr}} = \mathrm{MLP}_E(\mathbf{h}_i^{(T,0)}).
        \label{eq:site_energy}
    \end{equation}
\end{itemize}

Crucially, the charges $q_i$ and dipoles $\bm{\mu}_i$ are trained \textit{without direct
supervision}. No reference charges, dipole moments, or electrostatic potentials are
provided during training---only total energies and atomic forces. The learning signal
for these electrostatically meaningful quantities flows entirely through the long-range
energy term (Section~\ref{sec:pme}) and the total force
$\mathbf{F}_i = -\partial E_{\mathrm{total}}/\partial \mathbf{r}_i$, which couples
the spatial derivatives of $E^{\mathrm{lr}}$ back to the predicted charges and
dipoles via automatic differentiation. After training, the learned charges and
dipole-like vectors provide a decomposition of the electrostatic response. For
finite non-periodic structures, a molecular dipole can be constructed from both
charge positions and local dipoles,
$\mathbf{M}=\sum_i q_i\mathbf{r}_i+\sum_i\bm{\mu}_i$ after applying the fixed
sign convention of the dipole readout. For periodic condensed phases, the
corresponding polarization must be interpreted with the usual periodic-boundary
conventions rather than as a unique absolute dipole moment.

\subsection{Differentiable Particle-Mesh Ewald}\label{sec:pme}

We now present the central methodological contribution of this work: a fully
differentiable particle-mesh Ewald (PME) module that computes an FFT-accelerated
screened reciprocal-space electrostatic term from predicted atomic multipoles at
$O(N \log N)$ cost and supports end-to-end gradient backpropagation. By ``fully differentiable''
we mean that the entire computational graph---from the MPNN-predicted charges
and dipoles through discrete mesh assignment (indexed scatter), forward and
inverse real-to-complex FFTs, $k$-space Green's function multiplication,
influence function correction, and energy summation---is constructed from
native PyTorch operations whose analytical gradients compose correctly under
\texttt{autograd}, enabling stable training without surrogate loss functions
or gradient approximations. Unlike a standalone classical Ewald calculation,
the neural-potential module does not evaluate an explicit real-space
$\mathrm{erfc}(\alpha r)/r$ pair term. The mesh term represents the smooth
long-range channel, while the complementary short-range part is learned by the
local HotPP energy. For comparison and validation, our implementation also
provides a conventional Ewald summation baseline~\cite{ewald1921}, which we use
to assess PME accuracy in Section~\ref{sec:results}.

\subsubsection*{2.3.1 Mesh construction}

Given a periodic simulation cell defined by the $3\times 3$ matrix
$\mathbf{C}$ whose rows are the lattice vectors, the inverse cell is
$\mathbf{C}^{-1}$. We denote the three direct lattice vectors by
$\mathbf{a}_x,\mathbf{a}_y,\mathbf{a}_z$, i.e., the rows of $\mathbf{C}$. The
corresponding reciprocal basis vectors
$\mathbf{b}_x,\mathbf{b}_y,\mathbf{b}_z$ are the rows of
$2\pi\mathbf{C}^{-T}$, so that
$\mathbf{a}_\alpha\cdot\mathbf{b}_\beta=2\pi\delta_{\alpha\beta}$. The mesh
dimensions $N_x, N_y, N_z$ are determined from a target grid spacing $\Delta$:
\begin{equation}
    N_\alpha^{\mathrm{target}} = \left\lceil \frac{2\|\mathbf{a}_\alpha\|}{\Delta} + 1 \right\rceil,
    \qquad \alpha \in \{x, y, z\},
    \label{eq:mesh_target}
\end{equation}
where $\mathbf{a}_\alpha$ denotes the $\alpha$-th row of $\mathbf{C}$. We
provide three strategies for mapping the target dimensions to the actual mesh
size, each offering a different tradeoff between FFT efficiency and monotonic
convergence with respect to the grid spacing $\Delta$:

\begin{itemize}
    \item \textbf{Power-of-two}: $N_\alpha = 2^{\lceil \log_2(N_\alpha^{\mathrm{target}}) \rceil}$.
    Fastest FFT, but the energy is a step function of $\Delta$---small reductions in
    $\Delta$ may not change $N_\alpha$ at all, complicating convergence studies.

    \item \textbf{Smooth} (\textbf{default}): $N_\alpha$ is the smallest integer $\ge
    N_\alpha^{\mathrm{target}}$ whose prime factors belong to $\{2, 3, 5\}$.
    These numbers are nearly as FFT-friendly as powers of two while providing finer granularity and near-monotonic
    energy convergence.

    \item \textbf{Direct ceil}: $N_\alpha = \lceil N_\alpha^{\mathrm{target}} \rceil$.
    Fully monotonic in $\Delta$; FFT performance may degrade for unfavorable
    $N_\alpha$ with large prime factors.
\end{itemize}

The grid spacing $\Delta$ and mesh strategy are user-specified hyperparameters;
alternatively, the user may specify $[N_x, N_y, N_z]$ explicitly.

\subsubsection*{2.3.2 Hockney--Eastwood spline assignment}

We distribute point charges and dipoles onto the regular mesh using the
Hockney--Eastwood spline charge-assignment functions~\cite{hockney1988}
summarized by Deserno and Holm~\cite{deserno1998a,deserno1998b}. For an
assignment order $p$ (supporting
$p=1$ through $p=7$), the one-dimensional weight function $W^{(p)}(\xi)$ is a
compact piecewise polynomial of degree $p-1$. These are the spline
particle-mesh assignment weights, not the Lagrange interpolation weights used
with the unmodified continuum Green function in the original PME construction.

Given an atomic position $\mathbf{r}_i$ represented as a row vector, we first
compute the fractional coordinates relative to the mesh:
$\mathbf{s}_i = \mathbf{r}_i\mathbf{C}^{-1}$, then
$\mathbf{x}_i = (N_x s_i^x, N_y s_i^y, N_z s_i^z)$. For odd assignment order $p$,
the nearest mesh point is at $\mathbf{n}_i = \mathrm{round}(\mathbf{x}_i)$, and the
offset is $\bm{\xi}_i = \mathbf{x}_i - \mathbf{n}_i$. For even $p$, we use
$\mathbf{n}_i = \lfloor \mathbf{x}_i \rfloor$ with offset
$\bm{\xi}_i = \mathbf{x}_i - (\mathbf{n}_i + \tfrac{1}{2})$. The three-dimensional
assignment weight is the separable product:
\begin{equation}
    W_{ijk}(\mathbf{r}) = W^{(p)}\!\left(\xi^x - u_i\right)
    W^{(p)}\!\left(\xi^y - u_j\right)
    W^{(p)}\!\left(\xi^z - u_k\right),
    \label{eq:w3d}
\end{equation}
where $(u_i, u_j, u_k)$ enumerates the $p^3$ mesh points in the interpolation stencil
centered at $\mathbf{n}_i$.

For predicted charges and dipoles, the mesh stores a single effective charge
density. With the shorthand
$W_x=W^{(p)}(\xi_i^x-u)$, $W_y=W^{(p)}(\xi_i^y-v)$,
$W_z=W^{(p)}(\xi_i^z-w)$, and similarly
$W_x'=\partial W^{(p)}(\xi_i^x-u)/\partial \xi_i^x$, the value scattered from
atom $i$ to the stencil point $(n_i^x+u,n_i^y+v,n_i^z+w)$ is
\begin{align}
    \rho_{\mathrm{mesh}}(\mathbf{n})
    =
    \sum_{i=1}^{N}\sum_{(u,v,w)}
    \delta^{\mathrm{P}}_{\mathbf{n},\,\mathbf{n}_i+(u,v,w)}
    \Big[
        &q_i W_x W_y W_z \nonumber\\
        &+ N_x\bigl(\bm{\mu}_i\cdot\mathbf{C}^{-1}_{:x}\bigr) W_x' W_y W_z \nonumber\\
        &+ N_y\bigl(\bm{\mu}_i\cdot\mathbf{C}^{-1}_{:y}\bigr) W_x W_y' W_z \nonumber\\
        &+ N_z\bigl(\bm{\mu}_i\cdot\mathbf{C}^{-1}_{:z}\bigr) W_x W_y W_z'
    \Big],
    \label{eq:charge_dipole_assign}
\end{align}
where the inner sum runs over the $p$ stencil points along each dimension and
$\delta^{\mathrm{P}}$ is a periodic Kronecker delta on the mesh. Equivalently,
if a stencil point falls outside the mesh, its three indices are wrapped back
into the ranges $0,\ldots,N_x-1$, $0,\ldots,N_y-1$, and $0,\ldots,N_z-1$,
respectively. Here $\mathbf{C}^{-1}_{:\alpha}$
denotes the $\alpha$-th column of the inverse cell matrix. The first line is
the monopole assignment; the remaining three lines are the dipolar effective
charge density obtained by differentiating the same separable assignment kernel
along the three mesh directions.

\subsubsection*{2.3.3 Influence function deconvolution}

A subtle but consequential artifact of particle-mesh assignment is that the
mesh density is a smoothed version of the particle density. A point charge is
not placed on a single grid point; it is distributed over the nearby stencil by
the assignment weights $W^{(p)}$. In continuum language, this operation is a
convolution of the true particle density with the assignment kernel. Therefore,
before considering the additional aliasing errors introduced by sampling on a
finite grid, the leading reciprocal-space effect is simply a multiplication by
the Fourier transform of the assignment kernel:
\begin{equation}
    \tilde{\rho}_{\mathrm{mesh}}(\mathbf{k}) =
    \tilde{\rho}_{\mathrm{true}}(\mathbf{k}) \cdot \tilde{W}(\mathbf{k}),
    \label{eq:convolution}
\end{equation}
where $\tilde{W}(\mathbf{k})$ is the Fourier transform of the assignment function.
For the separable assignment kernel used in our implementation, the frequency
response is approximated in grid-index space as
\begin{equation}
    B(\mathbf{n}) =
    \mathrm{sinc}^p\!\left(\frac{n_x}{N_x}\right)
    \mathrm{sinc}^p\!\left(\frac{n_y}{N_y}\right)
    \mathrm{sinc}^p\!\left(\frac{n_z}{N_z}\right),
    \label{eq:Bk}
\end{equation}
where $\mathbf{n}=(n_x,n_y,n_z)$ are the discrete FFT frequency indices and
$\mathrm{sinc}(x)=\sin(\pi x)/(\pi x)$ with $\mathrm{sinc}(0)=1$. This separable
form is the standard sinc response associated with spline particle-mesh
assignment~\cite{essmann1995}. Our implementation uses the separable factor in
Eq.~\ref{eq:Bk} as a lightweight deconvolution correction for the leading mesh
assignment error.

This filtering modifies the high-frequency part of the reciprocal-space density
and produces systematic mesh-dependent errors. At this detailed level, the
correction is a simplified, P3M-inspired influence-correction strategy motivated
by particle-mesh methods~\cite{hockney1988,deserno1998a,deserno1998b}: we reduce
the leading assignment error by applying a deconvolution factor
$\mathcal{C}(\mathbf{n})$ to the Green's function in reciprocal space:
\begin{equation}
    \tilde{G}_{\mathrm{corrected}}(\mathbf{k}) =
    \tilde{G}(\mathbf{k}) \cdot \mathcal{C}(\mathbf{n}), \qquad
    \mathcal{C}(\mathbf{n}) = \frac{1}{|B(\mathbf{n})|^{2\gamma}},
    \label{eq:influence}
\end{equation}
where $\gamma$ is the implementation parameter \texttt{influence\_power}. The
default is $\gamma=1$, so the correction is $1/|B|^2$. We do not use the full optimal P3M influence function of
Hockney--Eastwood or Deserno--Holm, which contains aliasing sums and depends on
the chosen differentiation operator. Thus, the present method should not be
described as a full optimal P3M implementation; it is a differentiable PME solver
that borrows the P3M idea of assignment-kernel compensation through a simple
separable deconvolution factor matched to the spline assignment response.

\subsubsection*{2.3.4 Reciprocal-space energy}

The mesh density $\rho_{\mathrm{mesh}}(n_x, n_y, n_z)$ is transformed to reciprocal
space via a 3D real-to-complex FFT:
\begin{equation}
    \tilde{\rho}(k_x, k_y, k_z) = \mathrm{FFT}\!\left[\rho_{\mathrm{mesh}}\right],
    \label{eq:fft}
\end{equation}
where the discrete FFT frequency indices $\mathbf{n}$ are mapped to physical
reciprocal vectors by
$\mathbf{k}=n_x\mathbf{b}_x+n_y\mathbf{b}_y+n_z\mathbf{b}_z$ using the reciprocal
basis vectors defined above.

The reciprocal-space Green's function for the Gaussian-screened Coulomb interaction is:
\begin{equation}
    \tilde{G}(\mathbf{k}) =
    \begin{cases}
        0, & \mathbf{k} = \mathbf{0}, \\[4pt]
        \displaystyle \frac{4\pi}{|\mathbf{k}|^2}
        \exp\!\left(-\frac{|\mathbf{k}|^2}{4\alpha^2}\right), & \mathbf{k} \neq \mathbf{0},
    \end{cases}
    \label{eq:green}
\end{equation}
where $\alpha$ is the Ewald splitting parameter. Equation~(\ref{eq:green}) is the
Fourier transform of the smooth Ewald long-range kernel
$\mathrm{erf}(\alpha r)/r$. In a conventional Ewald decomposition this term is
combined with a real-space pair contribution
$\mathrm{erfc}(\alpha r)/r$ so that their sum recovers the full Coulomb
interaction. In the present neural potential, only the smooth mesh term is
evaluated explicitly; the complementary short-range part is absorbed into the
learned local energy. A smaller $\alpha$ gives a smoother mesh kernel but leaves
more short-range electrostatics to the MPNN, whereas a larger $\alpha$ places
more of the Coulomb interaction in the mesh channel and therefore requires a
finer grid for comparable discretization accuracy. In practice, $\alpha$ is
treated as a fixed hyperparameter.

The corrected electrostatic potential in reciprocal space is:
\begin{equation}
    \tilde{\phi}(\mathbf{k}) = \tilde{G}(\mathbf{k}) \cdot \mathcal{C}(\mathbf{n}) \cdot
    \tilde{\rho}(\mathbf{k}),
    \label{eq:potential_k}
\end{equation}
and the real-space potential is recovered via inverse FFT:
\begin{equation}
    \phi(\mathbf{r}_{n_x, n_y, n_z}) = \mathrm{IFFT}\!\left[\tilde{\phi}\right].
    \label{eq:ifft}
\end{equation}

The reciprocal-space electrostatic energy is then implemented as
\begin{equation}
    E^{\mathrm{recip}} = \frac{1}{2V} \sum_{n_x, n_y, n_z}
    \rho_{\mathrm{mesh}}(\mathbf{n}) \; \phi(\mathbf{n}),
    \label{eq:energy_recip}
\end{equation}
where $V = |\det(\mathbf{C})|$ is the cell volume, and the $\mathbf{k}=\mathbf{0}$
term is excluded (as its contribution is canceled by the neutralizing background
for charge-neutral systems).

\subsubsection*{2.3.5 Self-interaction correction}

The reciprocal-space energy includes a spurious self-interaction of each Gaussian
charge distribution with itself. For point charges, the self-energy correction is:
\begin{equation}
    E^{\mathrm{self}}_q = -\frac{\alpha}{\sqrt{\pi}} \sum_{i=1}^{N} q_i^2.
    \label{eq:self_q}
\end{equation}

For point dipoles, the analogous Gaussian self-interaction
yields~\cite{toukmaji2000}:
\begin{equation}
    E^{\mathrm{self}}_\mu = -\frac{2\alpha^3}{3\sqrt{\pi}}
    \sum_{i=1}^{N} |\bm{\mu}_i|^2.
    \label{eq:self_mu}
\end{equation}
Equations~(\ref{eq:self_q})--(\ref{eq:self_mu}) are the continuum-limit expressions;
they depend on the Ewald splitting parameter $\alpha$ and on the predicted
charges or dipoles.

The total long-range electrostatic energy is:
\begin{equation}
    E^{\mathrm{lr}} = E^{\mathrm{recip}} + E^{\mathrm{self}}_q + E^{\mathrm{self}}_\mu.
    \label{eq:E_lr_total}
\end{equation}

Every step in the PME pipeline---mesh assignment (with \texttt{index\_put\_} and
\texttt{accumulate=True}), forward and inverse FFTs, Green's function multiplication
in $k$-space, and energy summation---is implemented using native differentiable PyTorch
operations, enabling end-to-end gradient flow from the total energy to the network
parameters $\bm{\theta}$ through the entire reciprocal-space solver
\cite{paszke2019}.

\subsection{Total energy and training}\label{sec:training}

The total potential energy of the system is the sum of short-range and long-range
contributions:
\begin{equation}
    E_{\mathrm{total}} =
    \underbrace{\sum_{i=1}^{N} E_i^{\mathrm{sr}}}_{\text{short-range MPNN}}
    \;+\;
    \underbrace{E^{\mathrm{lr}}\bigl(\{q_i\}, \{\bm{\mu}_i\},
    \{\mathbf{r}_i\}, \mathbf{C}\bigr)}_{\text{differentiable PME}}.
    \label{eq:E_total}
\end{equation}

Atomic forces are obtained by analytical differentiation of the total energy:
\begin{equation}
    \mathbf{F}_j = -\frac{\partial E_{\mathrm{total}}}{\partial \mathbf{r}_j}
    = -\frac{\partial E^{\mathrm{sr}}}{\partial \mathbf{r}_j}
    - \frac{\partial E^{\mathrm{lr}}}{\partial \mathbf{r}_j},
    \label{eq:forces}
\end{equation}
where both terms are computed via automatic differentiation. The long-range force
term $\partial E^{\mathrm{lr}}/\partial \mathbf{r}_j$ receives contributions
through two channels: the explicit dependence of the mesh assignment on atomic
positions, and the implicit dependence through the position-sensitive charge and
dipole predictions $\partial q_i/\partial \mathbf{r}_j$ and
$\partial \bm{\mu}_i/\partial \mathbf{r}_j$, which are captured by the
computation graph.

The model is trained by minimizing a combined loss function:
\begin{equation}
    \mathcal{L} = \lambda_E \cdot
    \frac{1}{N_{\mathrm{batch}}} \sum_{b}
    \left(\frac{E_b^{\mathrm{pred}} - E_b^{\mathrm{ref}}}{N_b}\right)^2
    \;+\;
    \lambda_F \cdot
    \frac{1}{3\sum_b N_b} \sum_{b, i, \alpha}
    \left(F_{b,i,\alpha}^{\mathrm{pred}} - F_{b,i,\alpha}^{\mathrm{ref}}\right)^2,
    \label{eq:loss}
\end{equation}
where $b$ indexes structures in the batch, $N_b$ is the number of atoms in
structure $b$, and $\lambda_E$, $\lambda_F$ are user-defined weights. Optionally,
virial stress can be included as an additional loss term for training on condensed-phase
data.

Optimization uses the Adam or AdamW optimizer with grouped weight decay:
embedding and readout layers receive no weight decay, interaction weights
(tensor product coefficients) receive nonzero weight decay, and biases receive
none. A linear learning rate warmup over the first $N_{\mathrm{warmup}}$ steps
is followed by exponential decay or ReduceLROnPlateau scheduling.

\section{Results}\label{sec:results}

\subsection{Example 1: charged dimer}

We first examine the charged-dimer dataset as a controlled long-range molecular
test case. Following the charged molecular dimer example discussed by King
et al.~\cite{king2025charges}, we use molecular-pair id 0 from the
LODE/BioFragment molecular-dimer data~\cite{burns2017bfdb}. The system is a
neutral ion pair composed of a $\mathrm{C}_3\mathrm{N}_3\mathrm{H}_{10}^{+}$
cation and a $\mathrm{C}_2\mathrm{O}_2\mathrm{H}_3^{-}$ anion, giving 23 atoms
per configuration. The monomer geometries are kept fixed and only the
intermolecular separation is varied, so the dataset is essentially a charged
binding curve rather than a thermally sampled molecular trajectory. The
reference labels are HSE06 hybrid-DFT energies and forces with a many-body
dispersion correction. The local split used here contains 10 training
configurations with separations from 6.63 to 12.23 \AA{} and 3 validation
configurations from 13.09 to 15.00 \AA{}. This example is useful because the
electrostatic interaction is strong, the distinction between short-range and
long-range descriptions is easy to interpret, and the corresponding PME and
Ewald calculations can be compared directly over a broad parameter range.

All charged-dimer models use the same five-layer HotPP with cutoff
$5.0$ \AA{}, a polynomial cutoff with $p=5$, and a
Bessel-MLP radial layer with 8 basis functions and hidden widths $[32,32]$.
The long-range variants are named ``charge-only'' when
only the charge head is enabled and ``charge+dipole'' when both the charge and
dipole heads are enabled; each is run with either explicit Ewald summation or
differentiable PME. The latent electrostatic variables are trained only from
total energies and forces, without direct charge or dipole labels.

\begin{figure}[t]
\centering
\includegraphics[width=0.98\textwidth]{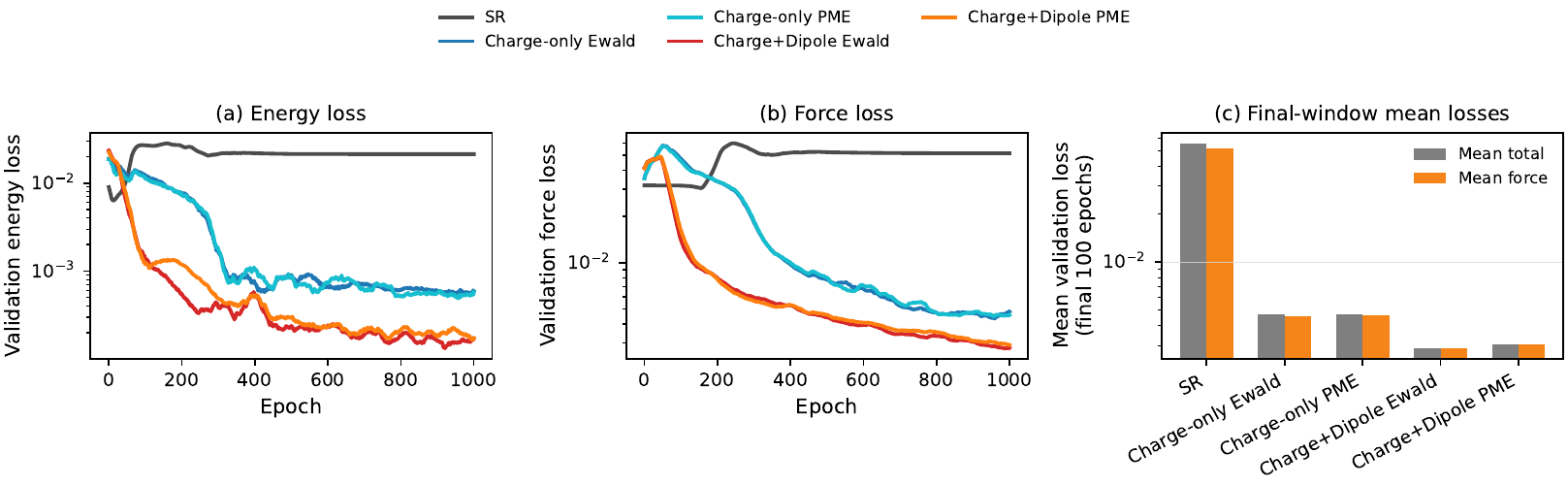}
\caption{Charged-dimer training behavior over the first 1000
epochs. (a) Validation energy loss and (b) validation force loss for the short-range baseline and the
long-range variants using either explicit Ewald summation or differentiable
PME. Because the raw epoch-wise losses oscillate strongly before full
convergence, panels (a) and (b) show 51-epoch centered moving averages in
log-loss space to emphasize the optimization trend. Panel (c) compares the
mean total and force validation losses over the final 100 epochs, computed
from the original epoch-wise values. The key trend is that long-range augmentation consistently
improves optimization, and that the charge+dipole models attain the lowest
final-window losses.}
\label{fig:charged_dimer_main}
\end{figure}

Figure~\ref{fig:charged_dimer_main} summarizes the first 1000 training epochs
for the tested models. Three observations are most relevant. First, the
long-range-augmented variants converge to
substantially lower energy and force validation losses than the short-range-only
baseline, showing that the charged-dimer interaction cannot be represented
efficiently by a purely local model. Second, the force loss is especially
sensitive to the long-range channel, consistent with the fact that forces probe
the derivative of the electrostatic interaction and therefore accentuate
deficiencies in a truncated local description. Third, the charge+dipole variants
reach the lowest final-window validation losses overall, indicating that a
dipolar response channel provides a meaningful correction beyond the leading
monopolar term even in this relatively simple test case.

\begin{figure}[t]
\centering
\includegraphics[width=0.98\textwidth]{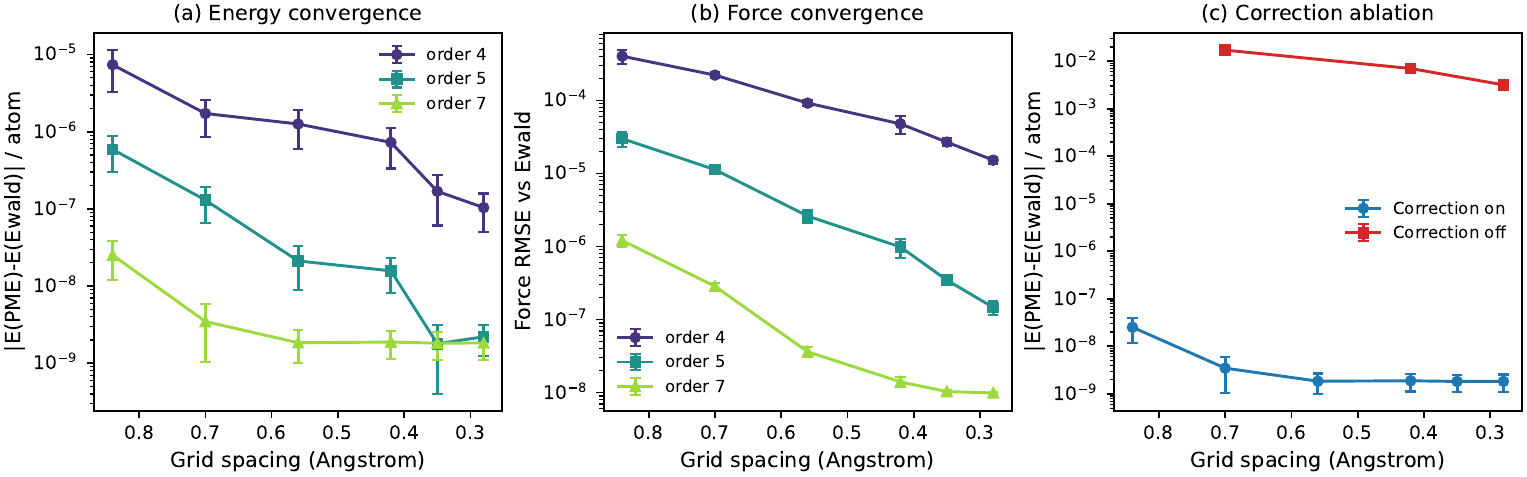}
\caption{PME--Ewald comparison on the charged-dimer test set.
(a) energy difference per atom and (b) force RMSE between PME and explicit
Ewald as a function of mesh spacing for several assignment orders.
(c) Ablation of the reciprocal-space influence correction at fixed order
$p=7$. The corrected PME converges systematically toward the Ewald reference,
while omitting the correction produces noticeably larger residual energy errors
at the same mesh resolution.}
\label{fig:charged_dimer_pme}
\end{figure}

An important methodological question is whether the differentiable PME module
reproduces the explicit Ewald reference with sufficient accuracy to be used as
a drop-in replacement during training and inference. Figure~\ref{fig:charged_dimer_pme}
addresses this by comparing PME against the direct Ewald calculation on the
same charged-dimer structures while varying the mesh spacing and interpolation
order. As expected, the PME error decreases systematically when the grid is
refined and when the assignment order is increased. The convergence is already
smooth at moderate interpolation order, and order $p=7$ gives the most stable
low-error regime across the tested grid spacings.

The correction ablation in Fig.~\ref{fig:charged_dimer_pme}(c) is particularly
important for the present implementation. Without the reciprocal-space
deconvolution factor, the mesh assignment smooths the charge density too
strongly, leaving a systematic bias in the reciprocal-space energy even when
the grid is moderately fine. Including the correction markedly reduces this
bias and restores the expected monotonic convergence toward the Ewald
reference. In practice, this means that physically reliable PME behavior can be
obtained on coarser meshes than would otherwise be required, which is exactly
the regime where PME becomes attractive as a scalable substitute for explicit
reciprocal-space summation.

Quantitatively, for assignment order $p=7$ with the correction enabled, the
mean PME--Ewald energy difference per atom is $3.47\times 10^{-9}$ eV at
grid spacing $0.70$ \AA{} and $1.83\times 10^{-9}$ eV at $0.28$ \AA{}; the
corresponding force RMSE decreases from $2.86\times 10^{-7}$ to
$9.91\times 10^{-9}$ eV/\AA{}. Disabling the correction at $0.28$ \AA{} raises
the energy difference per atom to $3.18\times 10^{-3}$ eV and the force RMSE to
$1.70\times 10^{-3}$ eV/\AA{}, demonstrating that the deconvolution correction
is essential for the reported mesh accuracy.

Taken together, the charged-dimer results establish two points that motivate
the rest of the paper. On the modeling side, they show that adding explicit
learned electrostatics improves both optimization and final validation behavior
relative to a purely local baseline, with the charge+dipole formulation being
the most effective variant among those tested. On the numerical side, they show
that the differentiable PME solver can reproduce direct Ewald energies and
forces to controlled accuracy, provided that a sufficiently smooth assignment
order and the reciprocal-space influence correction are used.

\subsection{Example 2: molten NaCl}

The molten-NaCl example tests whether the same long-range construction remains
useful in a dense ionic environment and whether the reciprocal solver retains
its expected scaling advantage beyond the small charged-dimer system. The
structures are derived from the liquid-NaCl dataset of Faller et
al.~\cite{faller2024density}, which contains 1014 configurations with 64 Na and
64 Cl atoms. In our converted split, 90\% configurations are used for
training and 10\% for validation.

The molten-NaCl models use a HotPP backbone with longrange method PME:
cutoff $6.0$ \AA{}, five interaction layers,
$n_{\mathrm{embedding}}=n_{\mathrm{hidden}}=64$, a
polynomial cutoff with $p=5$, and a Bessel-MLP radial layer with 8 basis
functions and hidden widths $[64,64]$. As above, the latent charges and
dipoles are not directly supervised. The replicated NaCl supercells used in the
timing study are generated only from the validation structures and are not used
as independent machine-learning generalization data.

\begin{figure}[H]
\centering
\includegraphics[width=0.98\textwidth]{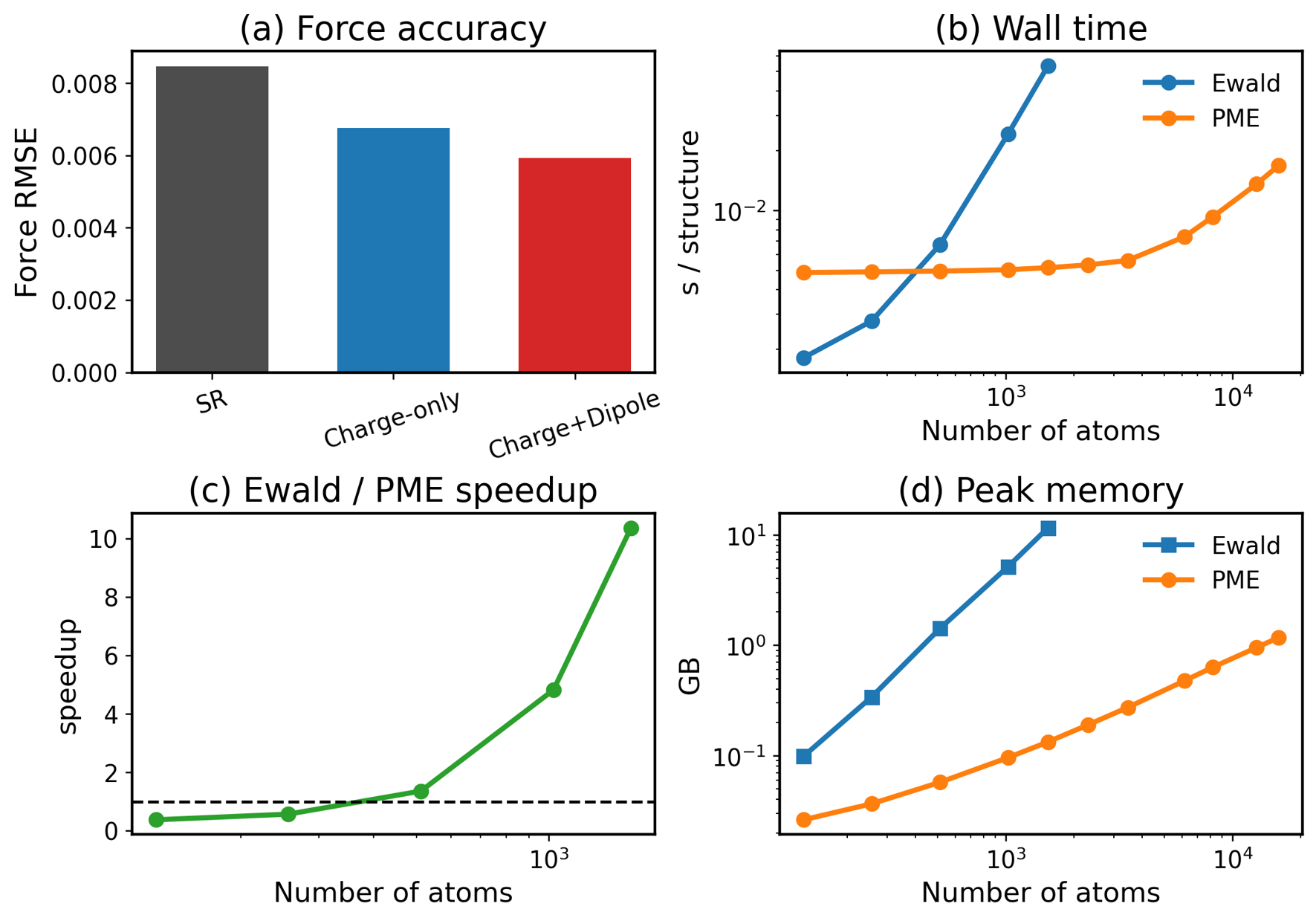}
\caption{Molten-NaCl accuracy and reciprocal-solver scaling summary. (a)
Force RMSE for the short-range, charge-only PME, and charge+dipole PME
models. (b) Median per-frame wall time for explicit Ewald and differentiable
PME on replicated molten-NaCl supercells. (c) Ewald/PME speedup on the
cells where explicit Ewald was run. (d) PME peak memory use, including
the 16,000-atom cell where explicit Ewald is intentionally omitted.}
\label{fig:molten_nacl_summary}
\end{figure}

We first examine the optimization behavior before comparing final RMSE values.
Both long-range variants reduce the validation total loss and force loss
relative to the short-range model, with the charge+dipole model giving the
lowest losses over the late training window. The final validation RMSE values
are summarized in Table~\ref{tab:molten_nacl_rmse}, and the force trend is
shown graphically in Fig.~\ref{fig:molten_nacl_summary}(a). The clearest
accuracy trend is in the forces: the force RMSE decreases from $8.46$
meV/\AA{} for the short-range baseline to $6.76$ meV/\AA{} with charge-only PME
and to $5.92$ meV/\AA{} with the charge+dipole PME model. Thus, the explicit
long-range channel reduces the force error by approximately 30\% relative to
the local baseline.

\begin{table}[H]
\centering
\caption{Molten-NaCl validation RMSE for the short-range baseline and the
PME-augmented long-range models. Energy errors are reported per atom.}
\label{tab:molten_nacl_rmse}
\begin{tabular}{@{}lcc@{}}
\toprule
Model & Energy (meV/atom) & Force (meV/\AA{}) \\
\midrule
Short-range & 0.573 & 8.46 \\
Charge-only PME & 0.763 & 6.76 \\
Charge+dipole PME & 0.845 & 5.92 \\
\bottomrule
\end{tabular}
\end{table}

The energy RMSE is slightly larger for the two long-range variants, increasing
from $0.573$ meV/atom to $0.763$ and $0.845$ meV/atom. These differences are
below $0.3$ meV/atom in absolute terms, and all three values remain in the
sub-meV/atom regime. We therefore regard the energy differences as effectively
at the numerical-error level for the present comparison, rather than as evidence
for a meaningful degradation of the potential. The force RMSE provides the more
diagnostic metric here, and it consistently improves when the explicit
long-range channel is included, as highlighted in
Fig.~\ref{fig:molten_nacl_summary}(a).

To isolate the numerical cost of the long-range solver, we also evaluated
molten-NaCl supercells from 128 to 16,000 atoms. These replicated structures are
used only as timing and numerical-consistency tests; they are not independent
machine-learning generalization data. For the supercells where explicit Ewald
was run, PME and Ewald agree to approximately $1.7$--$1.8\times 10^{-9}$
eV/atom in energy, with force-norm differences below $1.3\times 10^{-10}$ per
atom. The timing comparison in Fig.~\ref{fig:molten_nacl_summary}(b) shows the
expected crossover: explicit Ewald is faster at 128 atoms because the FFT
overhead dominates, while PME becomes faster at 512 atoms. The corresponding
Ewald/PME speedup in Fig.~\ref{fig:molten_nacl_summary}(c) reaches $4.82$ at
1024 atoms and $10.36$ at 1536 atoms. For larger cells, explicit Ewald was not
run because of its memory and time cost, while PME remains practical up to the
16,000-atom cell with a median frame time of $16.84$ ms and an average peak
memory footprint of $1.17$ GB, as shown in
Fig.~\ref{fig:molten_nacl_summary}(d).

These force-accuracy and scaling trends are collected in
Fig.~\ref{fig:molten_nacl_summary}, which keeps the RMSE comparison together
with the reciprocal-solver timing and memory data.

\section{Discussion}\label{sec:discussion}

We have introduced a differentiable particle-mesh Ewald framework for learning
long-range electrostatics within an E($n$)-equivariant Cartesian tensor message
passing potential. The model predicts atomic charges and dipoles from local
equivariant features, evaluates their long-range interaction through a fully
differentiable FFT-based solver, and combines this energy with the short-range
site energy predicted by HotPP. This construction keeps the favorable locality
and equivariance properties of modern MLIPs while restoring the non-local
electrostatic coupling required for ionic, polar, and interfacial systems.

The most important conceptual point is that the present work is not positioned
as a replacement for recent multipole-based Ewald MLIPs, but as a scalable
realization of the same emerging paradigm. Kim et al.~\cite{kim2026} have shown
that locally predicted multipoles can recover substantial long-range physics
from energy and force supervision alone. Our contribution is to make this
strategy compatible with the particle-mesh algorithms used in production MD,
including differentiable FFTs, spline assignment, and mesh correction. In this
sense, the method addresses a practical bottleneck rather than disputing the
physical premise of learned multipoles.

Several limitations should be acknowledged. The two example studies reported
here focus on monopoles and dipoles; quadrupolar channels are a natural next
extension of the Cartesian tensor representation but are not used in the
charged-dimer and molten-NaCl results above. The polarization response is
non-self-consistent: charges and dipoles are predicted in one forward pass from
local features, rather than iteratively relaxed in the field generated by all
other atoms. This choice keeps the model efficient and stable, but it may miss
feedback effects that are important in strongly polarizable materials. Finally,
as with all supervised MLIPs, transferability is bounded by the thermodynamic
states and chemical environments represented in the training data.

The combination of FFT-accelerated long-range electrostatics and end-to-end
differentiability opens a practical route to scalable simulations of polar
liquids, ionic solids, perovskites, solvated interfaces, and defect-containing
materials with learned long-range interactions. Future extensions will include
quadrupolar PME channels, explicit field-dependent response, and tighter
integration with active-learning workflows for generating long-range-sensitive
training data.

\section{Conclusions}\label{sec:conclusion}

We presented a differentiable particle-mesh long-range electrostatic formulation for equivariant neural network
potentials with learned atomic charges and dipoles. By replacing explicit
reciprocal-space Ewald summation with an FFT-based particle-mesh solver, the
framework targets $O(N\log N)$ evaluation of the smooth long-range channel while preserving
end-to-end gradient flow through the electrostatic energy, forces, and latent
multipole predictions. The Cartesian tensor representation of HotPP provides a
natural interface to this long-range module: scalar features predict charges,
vector features predict dipoles, and the short-range site energy absorbs the
local complementary part of the interaction.

The charged-dimer test case shows that learned long-range channels improve
optimization and that corrected PME can reproduce explicit Ewald energies and
forces on the same structures. The molten-NaCl example shows force-error
reduction from charge and dipole channels in a condensed ionic system and a
clear timing crossover where PME becomes substantially faster than explicit
Ewald summation. Together, these results support the central claim that
differentiable particle-mesh electrostatics can provide Ewald-level long-range
accuracy at particle-mesh cost within an equivariant neural-network potential.

\section*{Code availability}

The source code for the HotPP-LR implementation and the example workflows used
in this work is available in the \texttt{lr} branch of the HotPP GitLab
repository: \url{https://gitlab.com/bigd4/hotpp}.

\bibliographystyle{unsrt}

\end{document}